\documentclass[prc,aps,twocolumn]{revtex4}

\usepackage{graphicx}
\usepackage{color}
\usepackage{mathtools}
\begin{document}
%\draft
\title{ Properties of isospin asymmetric matter derived from chiral effective field theory.                                                    
} 

\author{Randy Millerson and Francesca Sammarruca}
\affiliation{Department of Physics, University of Idaho, Moscow, ID 83844, USA}

%\author{Francesca Sammarruca}
%\affiliation{Department of Physics, University of Idaho, Moscow, ID 83844, USA}

\begin{abstract}
We present and discuss properties of isospin asymmetric matter whose equation of state is derived from recent high-quality chiral nucleon-nucleon potentials and chiral effective three-nucleon forces. After a brief review of the chiral few-nucleon forces which we adopt, we concentrate on the symmetry energy and its density derivatives. We also explore the correlation between the symmetry energy at saturation density and its slope parameter, $L$. We estimate the truncation error across three orders of the chiral expansion for both the symmetry energy as a function of density and the slope parameter. Through an energy-density functional inspired by the liquid drop model, we establish a simple connection to finite nuclei. Specifically, we address the symmetry energy coefficient, the so-called reference (or equivalent) density, as well as the neutron skin thickness for $^{208}$Pb and $^{48}$Ca.   
\end{abstract}
\maketitle 
        
\section{Introduction} 
\label{Intro}

The main goal of contemporary nuclear theory is to provide microscopic, model independent predictions of nuclear systems.
For that purpose, accurate, theory-based few-nucleon forces must be developed and applied, 
consistently, in many-body systems.          

In the past, popular approaches to the construction of nuclear forces have been based on meson-theoretic or phenomenological models~\cite{CDbonn,Nij94,AV95}, implemented by 3NFs           
also derived from meson theory or phenomenology~\cite{TM,UIX}. More recently,              
chiral effective field theory ($\chi$EFT) has become popular in nuclear physics.
As in any effective field theory, one must define a ``resolution scale"            
and appropriate effective degrees of freedom, which, in $\chi$EFT, are pions and nucleons.   
In $\chi$EFT, two- and few-nucleon forces are derived systematically and on an equal footing~\cite{Wein91,Wein92,ME11}. Furthermore, at each order of the chiral expansion, it is possible to estimate a meaningful theoretical uncertainty.        

The purpose of this paper is to apply recently developed high-quality nucleon-nucleon (NN) chiral                        
 potentials~\cite{Ent2017} in order 
to explore properties of isospin asymmetric nuclear matter, specifically neutron-rich matter. 
The equations of state for symmetric and neutron matter based 
on Ref.~\cite{Ent2017} were recently presented in Ref.~\cite{Sam2018}.      

The interest in specific 
properties of the neutron-rich matter EoS originates from their relation to important (direct or 
indirect) observables.
In turn, measurements of those observables offer 
the opportunity to constrain aspects of the EoS.                                                          
Neutron skins are an examples of such sensitive ``observables", but unfortunately they are difficult to measure.
Experiments such as PREX II and CREX, however, are expected to measure the neutron radius of 
$^{208}$Pb and $^{48}$Ca with unprecedented accuracy~\cite{Mam2013}. 

Moreover, the EoS is a direct input in calculations of astrophysical systems, including supernovae and neutron stars (see, for instance, Ref.~\cite{Abb2017}). Microscopic calculations of compact astrophysical objects require knowledge of the EoS up to their central densities, which can be as large as 5-10 times normal density, and thus present enormous challenges. On the other hand, studies of the average-mass neutron star (as opposed to the maximum mass of a sequence) probe more moderate central densities. Recent calculations of our group based on $\chi$EFT find the radius of a 1.4 $M_{\odot}$ star to be in very good agreement with recent constraints~\cite{Sam2019}.                           

This paper is organized as follows. First, for the sake of completeness, we will briefly review some key elements of the two-nucleon forces (2NFs) and the three-nucleon forces (3NFs) which have been employed to develop the EoS applied here. We will then extract and discuss some isovector properties, that is, properties which arise from the isospin dependence of the EoS, paying particular attention to the order-by-order pattern of these quantities and their theoretical uncertainty. We will finish with a simple connection to finite nuclear systems established through a mass formula, and discuss our predictions of the asymmetry coefficient, the so-called ``equivalent density", and 
the neutron skins for some selected nuclei. 
     
\section{Brief review of theoretical tools}  
\label{II} 

We employ the EoS from microscopic calculations of nuclear and
neutron matter based on 2NFs and 3NFs as we will briefly outline below. 
The EoS is obtained as described in Ref.~\cite{Sam2018}. We apply the nonperturbative particle-particle
ladder approximation, corresponding to the leading-order contributions 
in the hole-line expansion of the energy per particle. 

\subsection{The two-nucleon forces}  
\label{IIA} 

The NN potentials
from Ref.~\cite{Ent2017} are available at                         
five orders of the $\chi$EFT expansion, 
from leading order (LO) to fifth order (N$^4$LO). These interactions are superior 
to the ones previously developed by the same group in that the same power counting scheme and 
regularization procedures are applied through all orders, thus ensuring better consistency. 
Furthermore, the long-range part of the interaction is fixed by the $\pi N$ LECs as determined in the 
recent and very accurate analysis of Ref.~\cite{Hofe+}. In fact, the                          
errors in the recently determined $\pi N$ LECs are small enough to be ignored for the purpose of uncertainty quantification.
At the fifth (and highest) order, the NN data below pion production threshold 
are reproduced with the excellent precision of $\chi ^2$/datum = 1.15.    

The necessary removal of high-momentum components prior to iteration of the potentials in the Lippmann-Schwinger
equation is accomplished through the application of the non-local regulator function:                              
\begin{equation}
f(p',p) = \exp[-(p'/\Lambda)^{2n} - (p/\Lambda)^{2n}] \,,
\label{reg_2NF}
\end{equation}
where $p' \equiv |{\vec p}\,'|$ and $p \equiv |\vec p \, |$ denote the final and initial nucleon momenta, respectively,  in the center-of-mass system.                     
The cutoff parameter, $\Lambda$, is taken to be smaller than or equal to 500 MeV, as those values have been associated 
with good perturbative behavior~\cite{Sam2018}.          

\subsection{The three-nucleon forces} 
\label{IIB} 

The leading 3NF appears at the third order of the chiral expansion 
(N$^2$LO). At this order, it consists of three contributions~\cite{Epe02}:
the long-range two-pion-exchange (2PE) term, 
the medium-range one-pion exchange (1PE) diagram, and a short-range contact term. 
We apply these 3NFs in the form of density-dependent effective two-nucleon interactions as derived in 
Refs.~\cite{holt09,holt10}, which facilitates their application in the particle-particle ladder approximation.

The effective density-dependent two-nucleon interactions consist of six one-loop topologies. Three of 
them are generated from the 2PE graph of the chiral 3NF and depend on the LECs
$c_{1,3,4}$, which are already present in the 2PE part of the NN interaction. 
Two one-loop diagrams are generated from the 1PE diagram and depend on 
the low-energy constant $c_D$. Finally, there is the one-loop diagram that involves the 3NF contact 
term, with LEC $c_E$.

The LECs $c_D$ and $c_E$ have been fixed within the three-nucleon sector, see Ref.~\cite{Sam2018} and references
therein for details. 
 The regulator function used in the 3NF               
 is the one of Ref.~\cite{Nav07},
i.e.\
\begin{equation}
  f(q)=\exp[(-q/\Lambda)^4]\ ,
  \label{eq:reg_3NF}
\end{equation}
where $q=| \vec p~'-\vec p \, |$ is the momentum transfer. With this choice, the 3NF is local
in coordinate space, making the construction of the $A=3$
wave functions less involved~\cite{Kie08}.

The complete 3NF beyond N$^2$LO is very complex and often neglected in nuclear structure
studies, but progress toward the inclusion of the subleading 3NF at N$^3$LO is underway 
\cite{Tew13,Dri16,DHS17,Heb15a}.
In the meantime, however, effectively complete
calculations up to N$^4$LO are possible for the 2PE 3NF. In Ref.~\cite{KGE12} it was
shown that the 2PE 3NF has essentially the same analytic structure at N$^2$LO, 
N$^3$LO, and N$^4$LO. Thus, one can add the three orders of 3NF contributions and 
parametrize the result in terms of effective LECs, which is the approach we are taking.   

Note that among all possible 3NF contributions, the 2PE 3NF was the first one to be 
calculated~\cite{FM57}. The prescription given above allows us to incorporate 
this most important 3NF contribution up to the highest orders considered~\cite{Sam2018}.

\begin{figure*}
    \centering
    \begin{minipage}{0.45\textwidth}
        \centering
        \includegraphics[width=0.9\textwidth]{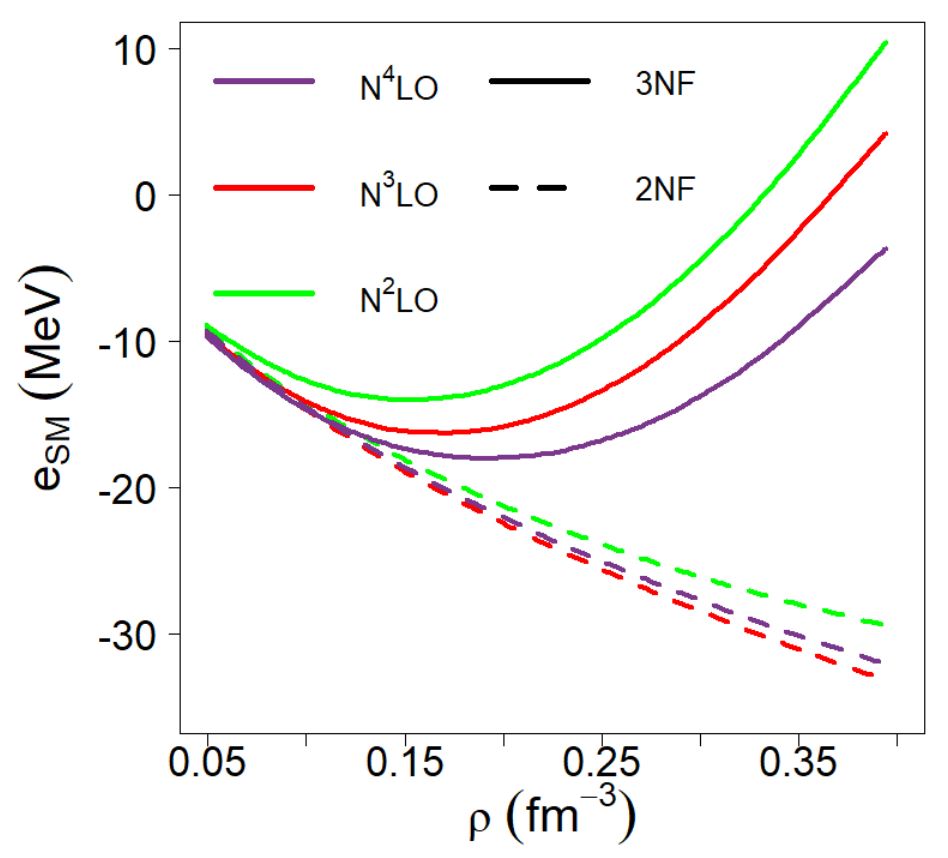}                             
        \caption{Energy per particle in symmetric nuclear matter as a function of density at three orders of chiral perturbation theory, with (solid lines) and without (dashed lines) 3NFs. The cutoff, $\Lambda$, is fixed at 450 MeV.}
    \label{esnm} 
    \end{minipage}\hfill
    \begin{minipage}{0.45\textwidth}
        \centering
        \includegraphics[width=0.9\textwidth]{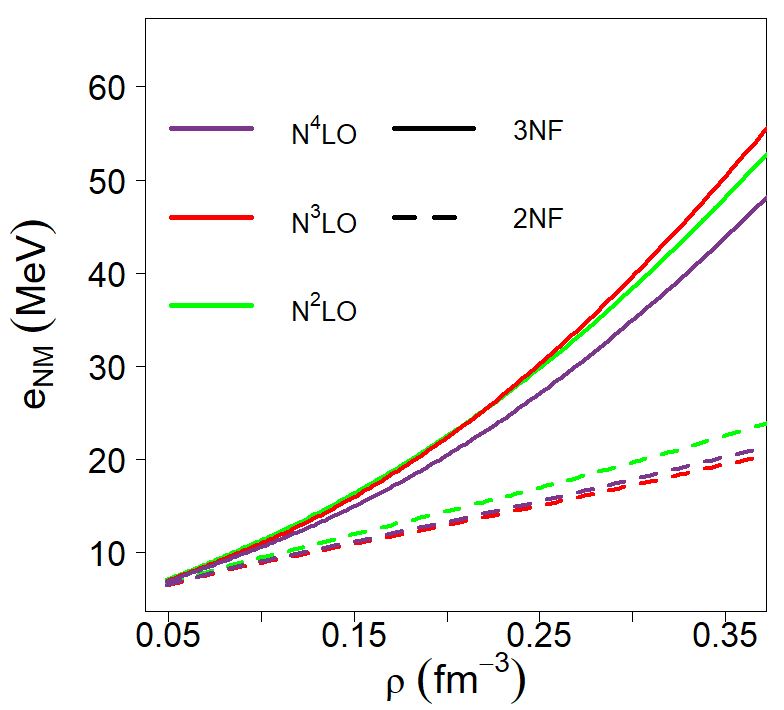}                          
        \caption{As in Fig.~\ref{esnm}, but for pure neutron matter.}
    \label{enm} 
    \end{minipage}
\end{figure*}

\begin{figure*}[t]
\centering
\includegraphics[width=7cm]{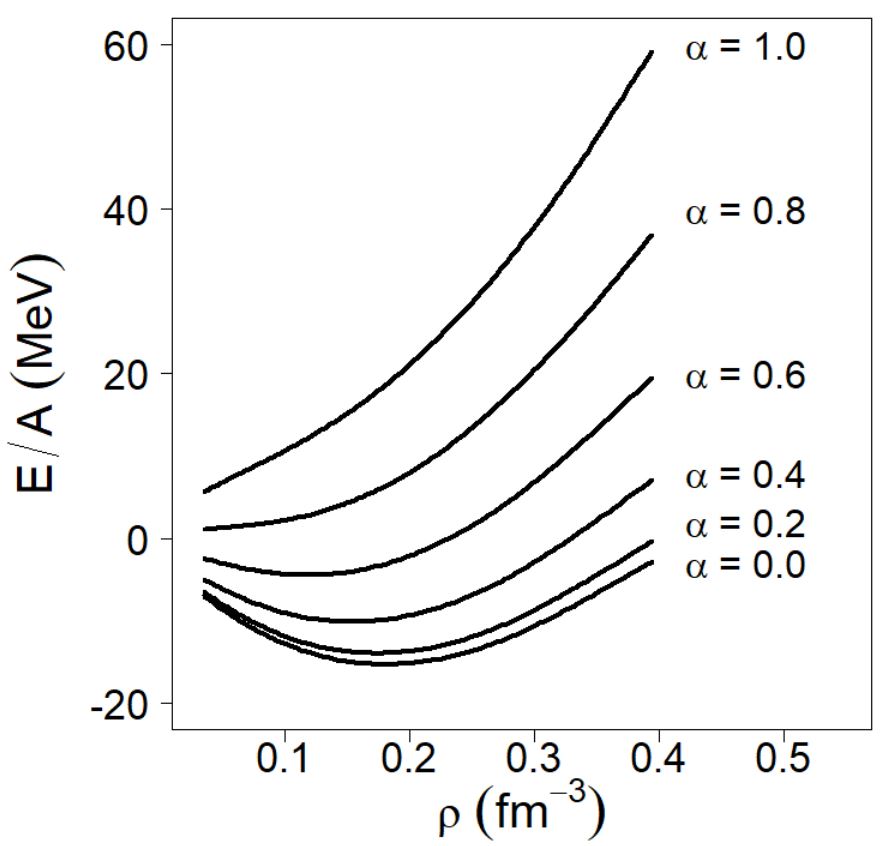}
\caption{Energy per particle in nuclear matter ranging from isospin symmetric to pure neutron matter. The predictions are obtained at N$^3$LO of the 2NF (and 3NFs as described in Sec.~\ref{IIB}) with a cutoff of 450 MeV.}
\label{e_asy}
\end{figure*}

\begin{figure*}[t]
\centering
\includegraphics[width=7cm]{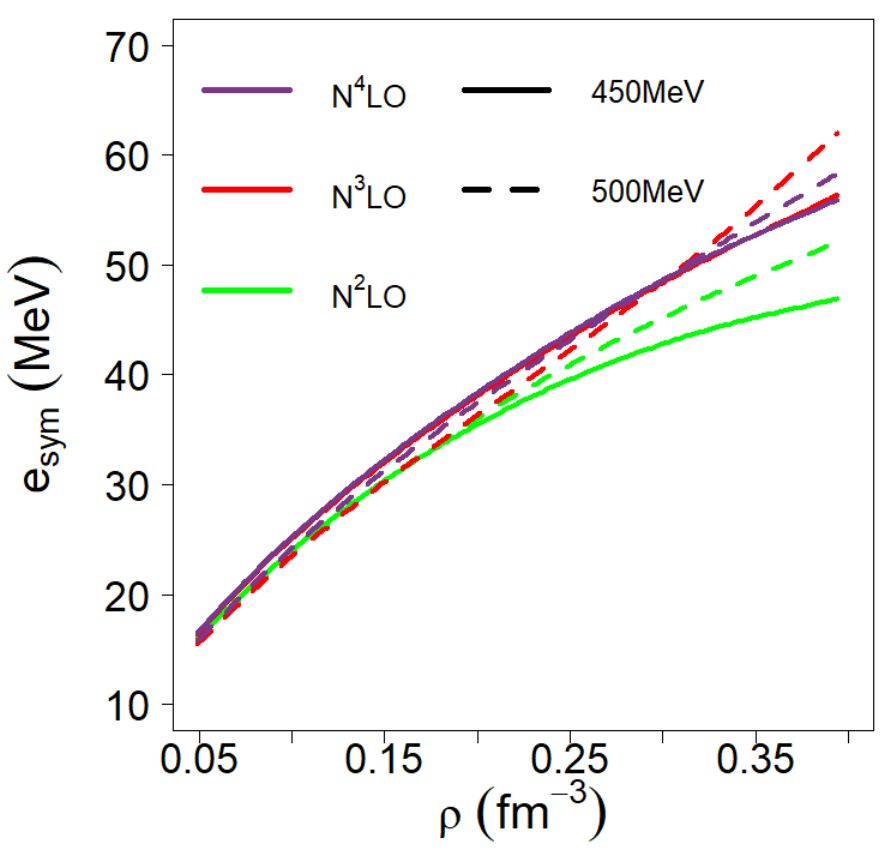}
\caption{Symmetry energy {\it vs.} density.}
\label{esym}
\end{figure*} 

\begin{figure*}[t]
\centering
\includegraphics[width=7cm]{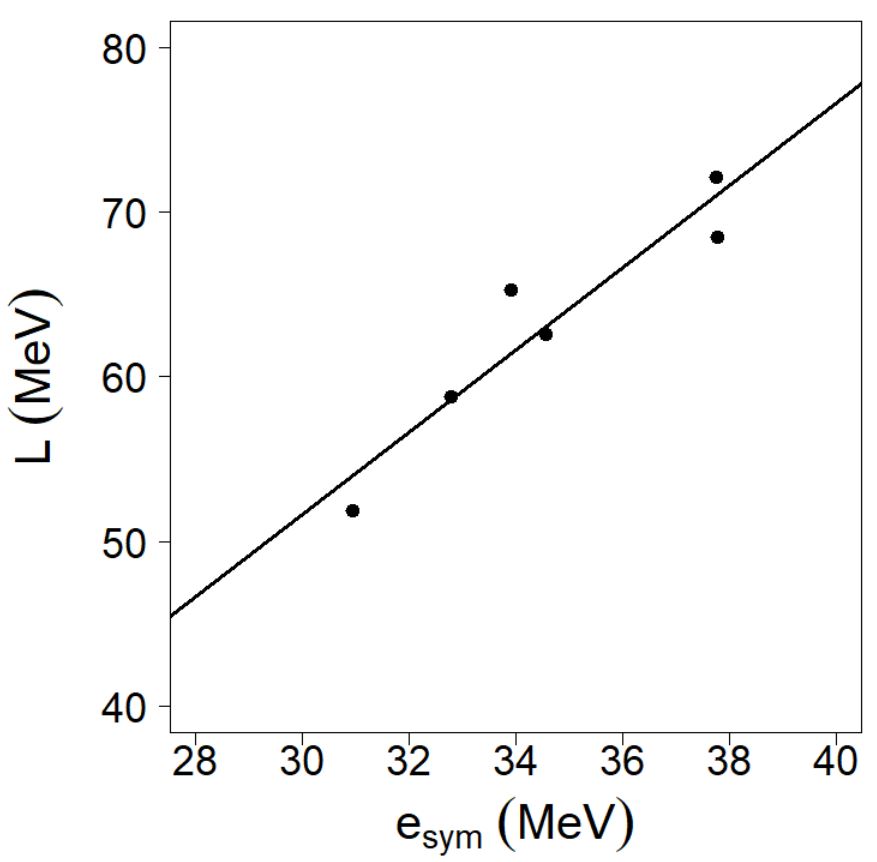}
\caption{Correlation between symmetry energy at saturation and the slope parameter, $L$. 
We find the correlation coefficient to be 0.945. }
\label{L_esym} 
\end{figure*}

\section{Some Properties of Isospin Asymmetric Matter} 
\label{III} 

The EoS of nuclear matter carries important information  about the nature of nuclear forces in many-body systems. From the structure of heavy isotopes to the mass-radius relation of neutron stars, the EoS of, in particular, neutron-rich matter plays a key role towards a better understanding of nuclear interactions and their density as well as isospin dependence. For these reasons, a large number of theoretical studies and experimental investigations have been and are being devoted to predicting or constraining EoS properties for both symmetric and asymmetric matter, see, for instance, Refs.~\cite{Sam2018,HS10,Heb11,Gez13,Baa13,Hag14b,Cor13,Cor14,Dri16,Tew16,Hol17,Bom2018,Oer2017,Bal2016,Fev2015,Sam2015,Hor2014,Ver2014,Tra2012,Ste2005,BL91,Sam2014}.                 

\subsection{Review of basic concepts and definitions}

Nuclear matter is an idealized infinite system of nucleons interacting {\it via} nuclear forces and is characterized, at zero temperature, by the energy per particle as a function of density.                                                                  
Isospin asymmetric nuclear matter refers to the presence of neutrons and protons in different concentrations, and thus is of relevance for neutron rich systems such as medium to heavy nuclei. The isospin asymmetry parameter measures the degree of isospin asymmetry and is defined as:
\begin{equation}
\alpha = \frac{(\rho_{n}-\rho_{p})}{(\rho_{n} + \rho_{p})} \; ,
\end{equation} 
where $\rho_n$ and $\rho_{p}$ are the neutron and proton densities, respectively. 

The energy per particle in nuclear matter at some density can be expressed as an expansion in terms of the asymmetry parameter:    
\begin{equation}
e(\rho, \alpha) =  e(\rho, \alpha = 0) + \frac{1}{2} \bigg( \frac{(\partial^{2} e(\rho,\alpha))}{\partial \alpha^{2}} \bigg)_{\alpha = 0} \alpha^{2} + \mathcal{O}(\alpha^4 ) \; .
\end{equation}
The expansion above is usually approximated by the well-known parabolic form:                                  
\begin{equation}
e(\rho, \alpha) \approx e_0 (\rho) + e_{sym}(\rho) \ \alpha^2 \; , 
\end{equation}
where 
$ e_0 (\rho) = e(\rho, \alpha=0)$.                 
Note that the validity of the parabolic approximation has been verified microscopically up to fairly high densities~\cite{BL91}.                                                                                                                                                               
Within this approximation, the symmetry energy, $e_{sym}$, is the difference between the energy per particle in neutron matter and the one in symmetric matter. An expansion of the symmetry energy about the saturation density, $\rho_0$, yields:                                                                           
\begin{equation}
\label{esym_exp} 
e_{sym}(\rho) \approx e_{sym} (\rho_0) + \frac{L}{3} \frac{\rho - \rho_0}{\rho_0} + \frac{ K_{sym}}{18} \Big (\frac{\rho - \rho_0}{\rho_0}\Big )^2 \; + ... \;.
\end{equation}  

The slope parameter, $L$, is of particular interest, 
 being a good measure for the density dependence of the symmetry energy at normal density: 
\begin{equation}
L = 3 \rho_0 \bigg( \frac{\partial e_{sym} (\rho)}{\partial \rho} \bigg)_{\rho_0} \; . 
\end{equation}
In fact, the slope parameter of the symmetry energy is a remarkable quantity because of its correlation with the neutron skin thickness found in neutron rich nuclei. Ongoing or planned experiments with electroweak probes seeking to measure the neutron radius of $^{208}$Pb and $^{48}$Ca, such as PREX II and CREX, respectively, highlight the importance and timeliness of theoretical investigations. Such experiments promise to provide accurate measurements of the neutron skin, and hence reliable constraints on the symmetry pressure, to which the slope parameter is closely related. Moreover, the radius of the typical-mass neutron star has been found to be sensitive to the pressure in neutron matter at normal density, see Ref.~\cite{Sam2019} and the comprehensive list of citations therein.

The curvature parameter or isovector incompressibility, $K_{sym}$, is associated with the next higher-order derivative of the symmetry energy at saturation density and is defined as:          
\begin{equation}
K_{sym} =                                                     
 \ 9 \rho_0^2 \bigg( \frac{\partial^2 e_{sym} (\rho)}{\partial \rho^2} \bigg)_{\rho_0} \; . 
\end{equation} 
Linear correlations have been explored between both $L$ and $K_{sym}$ and the symmetry energy at saturation, $e_{sym} (\rho_0)$~\cite{San2015, Ton2018, Hol2018}. The theoretical uncertainty for $K_{sym}$ is large and due in part to the 
corresponding uncertainty on the isoscalar incompressibility. Efforts to constrain the symmetry energy curvature have 
encountered considerable challenges~\cite{Vid09,Duc11,Santos14}.
We will not focus on incompressibility correlations at this time.                                                                         

\begin{table*}
  \caption{ Predicted values of symmetry energy and related properties at three orders of chiral perturbation theory and two values of the cutoff parameter obtained as 
explained in the text. 
}
\label{isovec_vals}
\centering
\begin{tabular*}{\textwidth}{@{\extracolsep{\fill}}cccccccc}
\hline
\hline
   & $\Lambda$ (MeV) & $e_{sym} (\rho_0)$ (MeV) & $L$ (MeV) & $K_{sym}$ (MeV)\\
\hline     
\hline
N$^2$LO & 450 & $30.9 \pm 3.6$ & $51.9 \pm 10.7$ & $-93.2 \pm 27.9$\\
        & 500 & $32.8 \pm 1.1$ & $58.8 \pm 6.4$  & $-85.0 \pm 43.0$\\
\hline 
N$^3$LO & 450 & $34.6 \pm 3.2$ & $62.6 \pm 5.9$ & $-65.3 \pm 17.9$\\
        & 500 & $33.9 \pm 3.9$ & $65.2 \pm 6.9$  & $-42.0 \pm 5.9$\\
\hline
N$^4$LO & 450 & $37.8 \pm 1.5$ & $68.5 \pm 3.0$ & $-83.2 \pm 8.4$\\
        & 500 & $37.8 \pm 1.7$ & $72.1 \pm 2.8$  & $-47.9 \pm 2.5$\\
\hline
\hline
\end{tabular*}
\end{table*}

\subsection{Order-by-order predictions and error estimates} 

We begin by showing the various EoS which we use in the present work.
Given that the first order at which a realistic NN interaction can be constructed is the third order (N$^2$LO), 
we will present predictions at third and higher orders. 
Those are displayed in Fig.~\ref{esnm} and Fig.~\ref{enm} for symmetric and pure neutron matter, respectively. For this
display, the cutoff has been fixed at 450 MeV. 
As is well known, the inclusion of 3NFs is essential to develop an EoS for symmetric nuclear matter with realistic saturating behavior, see Fig.~\ref{esnm}. The effect of including 3NFs is strongly density dependent. Figure~\ref{enm} shows that the effects of 3NFs  is also strong in neutron matter, although to a somewhat lesser degree.

Figure~\ref{e_asy} displays the energy per particle in isospin asymmetric matter as a function of density and for increasing degree of asymmetry, cf. Eq.~(5), for one selected order and cutoff. 

As this paper concentrates on isovector properties of the EoS, the spot light now moves onto the symmetry energy, which is shown in Fig.~\ref{esym} at the three chiral orders and both cutoff values. 
We will proceed treating the six cases shown in Fig.~\ref{esym} as six realistic interactions in their own right. 

Table I displays values for the parameters defined in the previous section, see Eq.~(\ref{esym_exp}), 
for the six EoS under consideration. From the Table, we observe that the range of $L$ across all chiral orders can be stated as $58.1 \pm 16.8$ MeV. The symmetry energy at saturation is within the range of $33.4 \pm 6.1$ MeV. Note 
that the actual saturation density for each interaction is used. As apparent from Fig.~\ref{esnm}, the latter covers a 
range of approximately 0.16-0.20 fm$^{-3}$.

Figure~\ref{L_esym} displays the slope parameter {\it vs.} the symmetry energy at saturation density. The relation is approximately linear with a correlation of 0.945. Here the standard Pearson correlation coefficient is used, which is defined as:      
\begin{equation}
P(x,y) = \frac{cov(x,y)}{\sigma_x \sigma_y} \; ,
\end{equation}
where $\sigma$ is the standard deviation and $cov(x,y)$ is the covariance:
\begin{equation}
cov(x,y) = \sum^n_i \frac{(x_i - \bar{x})(y_i - \bar{y})}{n-1} \; .
\end{equation}  

One of the strengths of $\chi$EFT is the opportunity of order-by-order improvement. At each order, the truncation error should be a reasonable measure of the uncertainty arising from omitting the next order contributions.    
The truncation error is computed based upon the level of knowledge of an observable at some given chiral order. Therefore, if the observable's value, $X$, is known, say, at order $n+1$, than the truncation error at order $n$ can be estimated as the difference between the two values at order $n+1$ and $n$:
\begin{equation}
\epsilon_n = | X_{n+1} - X_n | \; .
\end{equation}
This is, indeed, a reasonable measure of what one is missing by retaining only terms up to order $n$. 
However, if $X_{n+1}$ is not known, than we use the prescription~\cite{Ent2017}:                
\begin{equation}
\epsilon_n \approx |X_n - X_{n-1}| \frac{Q}{\Lambda} \; ,
\label{eps} 
\end{equation}  
where $Q$ is the typical momentum of the system and $\Lambda$ is the momentum cutoff.
For the fifth (and highest) order, we use Eq.~(\ref{eps}) taking $Q$ to be 
the r.m.s. value of the relative momentum of two neutrons in neutron matter at the specified density, which can be 
estimated to be about 60\% of the Fermi momentum~(see Ref.~\cite{Sam2019} and references therein).
(Given that the highest momentum in neutron-rich matter is smaller than the one in pure neutron matter, our 
choice is likely to overestimate, rather than underestimate, the truncation uncertainty.)

To state our final results at N$^3$LO for the symmetry energy, the slope parameter, and the isovector incompressibility,
we apply the following procedure.
We average the predictions of the observable $X$ for the two values of the cutoff separately at N$^3$LO and N$^4$LO,
yielding $\bar{X}_4$ and $\bar{X}_5$, respectively. The truncation error at 
N$^3$LO is then estimated to be $\Delta_X = |\bar{X}_4 - \bar{X}_5|$.                      
Alternatively, one can take the largest of the errors at the two cutoff values, which is the value of
$\Delta_X$ given in parentheses.

For the symmetry energy, the slope parameter, and the isovector incompressibility at N$^3$LO we obtain
(all numbers in MeV):                                             
\begin{equation}
e_{sym} = 34.3 \pm \Delta_{e_{sym}} \; \; \; \; \; \; \Delta_{e_{sym}} = 3.6 (3.9) \; , 
\label{err1}
\end{equation}
\begin{equation}
L = 63.9 \pm \Delta_L \; \; \; \; \; \; \Delta_L = 6.4 (6.9) \; , 
\label{err2}
\end{equation}
\begin{equation}
K_{sym} = -53.7 \pm \Delta_{K_{sym}} \; \; \; \; \; \; \Delta_{K_{sym}} = 11.9 (17.9)       \; . 
\label{err3}
\end{equation}
We see that $K_{sym}$ shows large variations, which reflect the extreme sensitivity of the second derivative to the details of each of the curves in Fig.~\ref{esym}. \\
A phenomenological study of the EoS based on Skyrme density functionals~\cite{Ala2014} reports    
the isovector incompressibility to be within the range $K_{sym} = -22.9 \pm 73.2$ MeV, whereas the slope parameter is stated as $L = 65.4 \pm 13.5$ MeV.

\section{Connecting with nuclei} 
\label{VI} 

\subsection{Brief review of concepts}

Using the energy per particle in infinite matter as given in Eq.(5), we can establish a simple but direct connection with the energy per nucleon in a spherically symmetric nucleus through the semi-empirical mass formula:
\begin{equation}
E(Z,A) = \int d^3 r \ \rho(r) \ e(\rho,\alpha) + \int d^3 r \ f_0 \ |\nabla \rho|^2 + E_{Coul} \; .
\end{equation}
where the Coulomb contribution is written as:      
\begin{equation}
E_{Coul} = \frac{e^2}{\epsilon_0} \int^{\infty}_{0} d r^{'} [r^{'} \rho_{p}(r^{'}) \int^{r^{'}}_{0} d^3 r \ \rho_{r}(r) ] \; .
\end{equation}  

Note that the surface term in the equation above does not include a contribution from the isovector density, 
$\rho_n - \rho_p$, as such effects were demonstrated to be negligible~\cite{Fur2002}. The parameter $f_0$ is a fitted constant for which we used a value of $65 \ MeV \ fm^{5}$, consistent with the range determined in Ref.~\cite{Oya2010}. 

We use the two-parameter Thomas-Fermi distribution function to describe the nucleon density as a function of the radial coordinate from the center of the nucleus:                                
\begin{equation}
\rho (r) = \frac{\rho_{a}}{1 + e^{(r - r_b)/c}} \; .
\end{equation}
The parameters $r_b$ and $c$, radius and diffuseness, respectively, are themselves evaluated through minimization of the energy per nucleon.                                                                                                                                                                        

\subsection{Symmetry energy coefficient and reference density}

\begin{figure*}
    \centering
    \begin{minipage}{0.45\textwidth}
        \centering
        \includegraphics[width=0.9\textwidth]{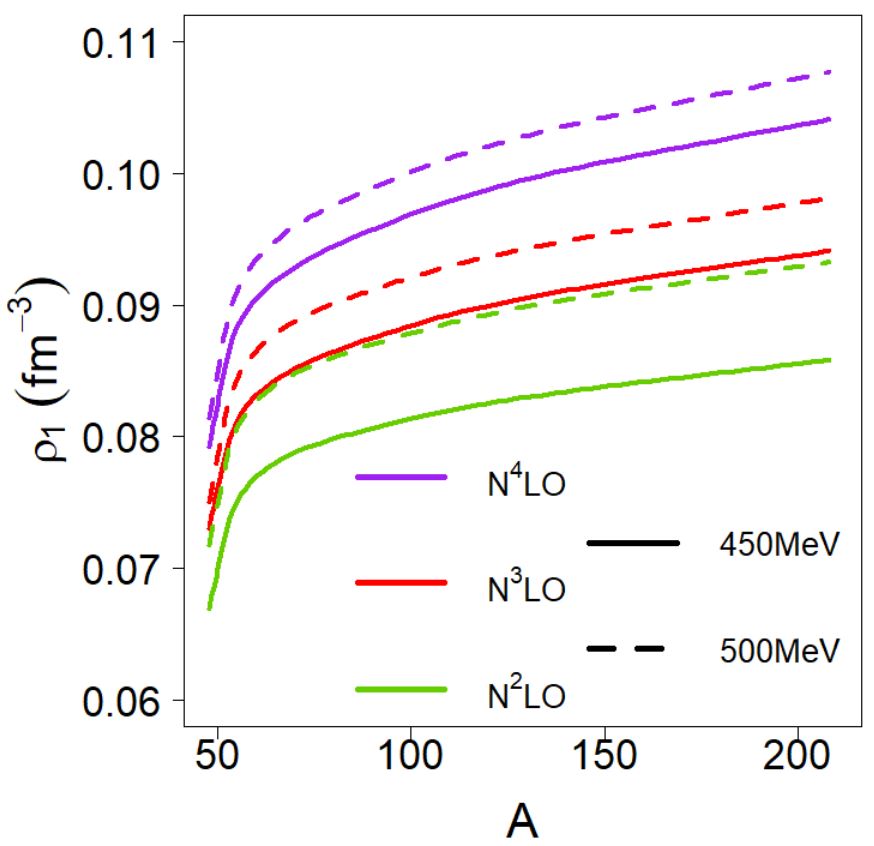}                         
        \caption{Reference density {\it vs.} atomic number.}
    \label{rhoref_A}
    \end{minipage}\hfill
    \begin{minipage}{0.45\textwidth}
        \centering
        \includegraphics[width=0.9\textwidth]{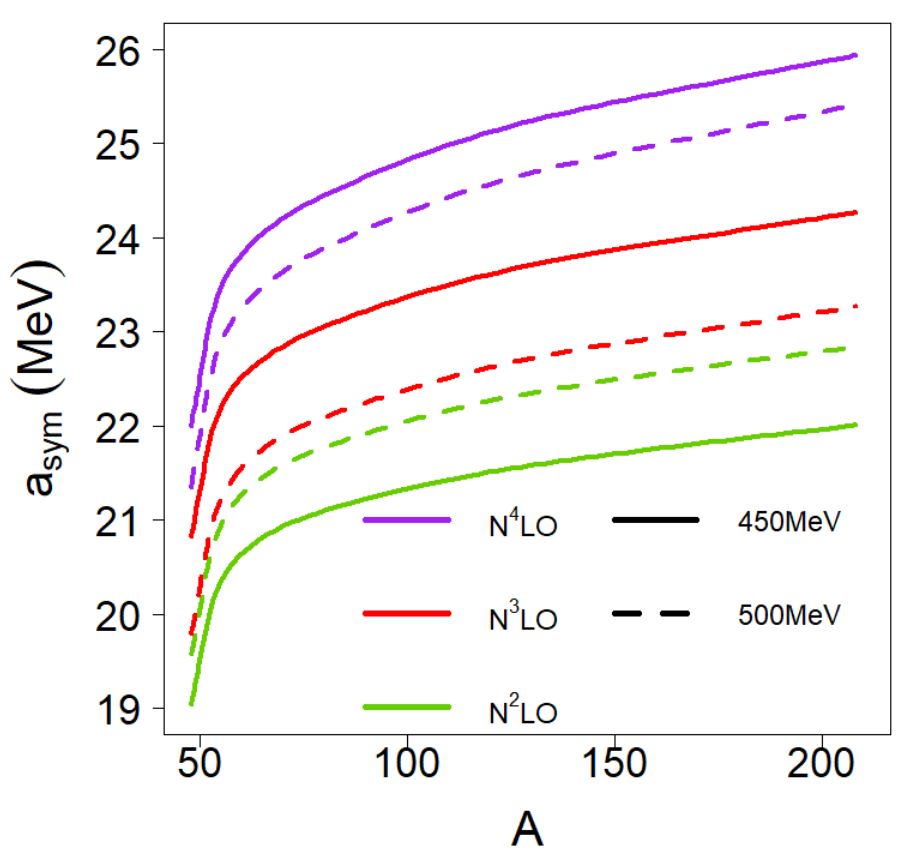}                         
        \caption{Symmetry energy coefficient {\it vs.} atomic number.}
    \label{coeff_A}  
    \end{minipage}
\end{figure*}

Having obtained density functions for neutrons and protons, we now proceed to calculate the symmetry energy coefficient in a nucleus, $a_{sym}$, which we write as                                                                    
\begin{equation}
a_{sym} (A,N) = \frac{A}{(N-Z)^2} \int \rho(r) \ e_{sym} (\rho) \ \alpha^2(r) \ d^3 r \; .
\label{asym} 
\end{equation}                                   

The reference density (say, $\rho_1$) for a particular nucleus (A,N) is defined as the density for which the symmetry energy is equal to the symmetry energy coefficient:
\begin{equation}
e_{sym} (\rho_1) = a_{sym} (A,N) \; .
\label{refdens}
\end{equation}
As actual measurements can only be performed with finite nuclei, clearly the reference density is a potentially useful quantity, which presents the opportunity to constrain the symmetry energy at the average density of the specific nucleus under consideration.

We now turn our attention to reference density predictions for a variety of neutron-rich nuclei, some of which are currently under extensive investigations, using the equation of states described in the preceding sections. As previously mentioned, the CREX and PREX II experiments will address specifically the neutron densities/radii/skins of $^{208}$Pb and $^{48}$Ca. In addition, we include $^{56}$Fe and $^{116}$Sb in our set of selected nuclei. We will observe how the reference density is impacted by the atomic number, while exploring its variations as we move across the chiral orders.
                              
Symmetry energy coefficients were extracted in Ref.~\cite{Liu2010} based on the liquid drop mass formula over a broad range of ``A" using more than 2000 precisely measured nuclear masses. Assuming a form 
$a_{sym}= S_0(1 + \kappa A^{-1/3})^{-1}$, where $S_0$ is the volume symmetry energy coefficient and $\kappa$ is the
ratio of the surface to the volume symmetry energy coefficients, a two-parameter fit resulted into 
$S_0 = 31.1 \pm 1.7$ MeV and $\kappa = 2.31 \pm 0.38$. Exploting the relation between the symmetry energy at reference
density and the symmetry energy coefficient of nuclei, Eq.~(\ref{refdens}), a range of 0.7 $\pm $ 0.1 was determined 
for the exponent $\gamma$ introduced through the parametrization $e_{sym}(\rho) = S_0 (\rho/\rho_0)^{\gamma}$. 
At the same time, the mass dependence of the reference density, which is a function of both $\kappa$ and $\gamma$, 
could be deduced, through the relation $ (\rho_1/\rho_0)^{\gamma} = 
(1 + \kappa A^{-1/3})^{-1}$.                                                                               
The range of the slope parameter, $L$, was then determined to be between 53 and 79 MeV~\cite{Liu2010}. 

Fig.~\ref{rhoref_A} shows our predicted reference density as a function of atomic number, while Fig.~\ref{coeff_A} displays the symmetry energy coefficient also as a function of the atomic number. We note that, for increasing chiral order, the reference density of $^{208}$Pb approaches the value of 0.1 $fm^{-3}$, which is consistent with the approximate reference density of $^{208}$Pb as determined in Ref.~\cite{Cen2009}. We found that the reference density for $^{208}$Pb has a range of $0.086 \ fm^{-3} \ \leq \rho_1 \leq 0.108  \ fm^{-3}$ across all orders and cutoffs, while the symmetry energy coefficient has a range of $19.1 $  MeV $ \leq a_{sym} \leq 25.9 $ MeV. 

With regard to our predicted reference density as a function of the atomic number, Fig.~\ref{rhoref_A}, we observe a similar behavior for all chiral orders and both cutoffs, where the values decrease rapidly for A less than 50 and approach a constant value for heavy nuclei, as it might be expected. We note, however, that the reference density vary considerably with chiral order, especially for large atomic numbers.

\subsection{Symmetry energy and neutron skins}

\begin{table*}
  \caption{Predicted neutron skin of $^{208}$Pb.                            
}
\label{ns_vals}
\centering
\begin{tabular*}{\textwidth}{@{\extracolsep{\fill}}ccccc}
\hline
\hline
  Order & $\Lambda$ = 450 MeV & $\Lambda$ = 500 MeV\\
\hline     
\hline
N2LO & 0.133 $\pm $ 0.010 fm & 0.140 $\pm $ 0.005 fm \\
N3LO & 0.143 $\pm $ 0.007 fm & 0.145 $\pm $ 0.011 fm \\
N4LO & 0.150 $\pm $ 0.004 fm & 0.156 $\pm $ 0.006 fm \\
\hline
\hline
\end{tabular*}
\end{table*}
\begin{table*}
  \caption{As in Table~\ref{ns_vals} but for $^{48}$Ca.                            
}
\label{ns_ca_vals}
\centering
\begin{tabular*}{\textwidth}{@{\extracolsep{\fill}}ccccc}
\hline
\hline
  Order & $\Lambda$ = 450 MeV & $\Lambda$ = 500 MeV\\
\hline     
\hline
N2LO & 0.132 $\pm $ 0.011 fm & 0.134 $\pm $ 0.002 fm \\
N3LO & 0.143 $\pm $ 0.005 fm & 0.136 $\pm $ 0.006 fm \\
N4LO & 0.138 $\pm $ 0.002 fm & 0.142 $\pm $ 0.002 fm \\
\hline
\hline
\end{tabular*}
\end{table*}

We now move to neutron skins, specifically of $^{208}$Pb and $^{48}$Ca, as predicted by the EoS described in the previous sections. As mentioned before, the neutron skin thickness, particularly for $^{208}$Pb, is of great contemporary interest due to its close relation to the slope of the symmetry energy, and therefore has been studied extensively, for instance, Refs.\cite{Sam22018,Sam2016,Ina2015,Roc2011,War2009,Xu2011,Roc2013}. 

The neutron skin is defined as the difference between the root-mean-squared (r.m.s.) radii of the neutron and proton density distributions:             
\begin{equation}
S_n = R_n - R_p \; ,
\end{equation}
where 
\begin{equation}
R_i = \sqrt{\frac{1}{T_i} \int^{\infty}_{0} \rho_{i} (r) \ r^2 \ d^3r} \; ,
\end{equation}  
$i=n,p$ and $T_n, \ T_p = N, \ Z$ respectively. 
We calculate neutron skins using the six chiral interactions considered previously.                                                    
\begin{figure}[t]
\centering
\includegraphics[width=6cm]{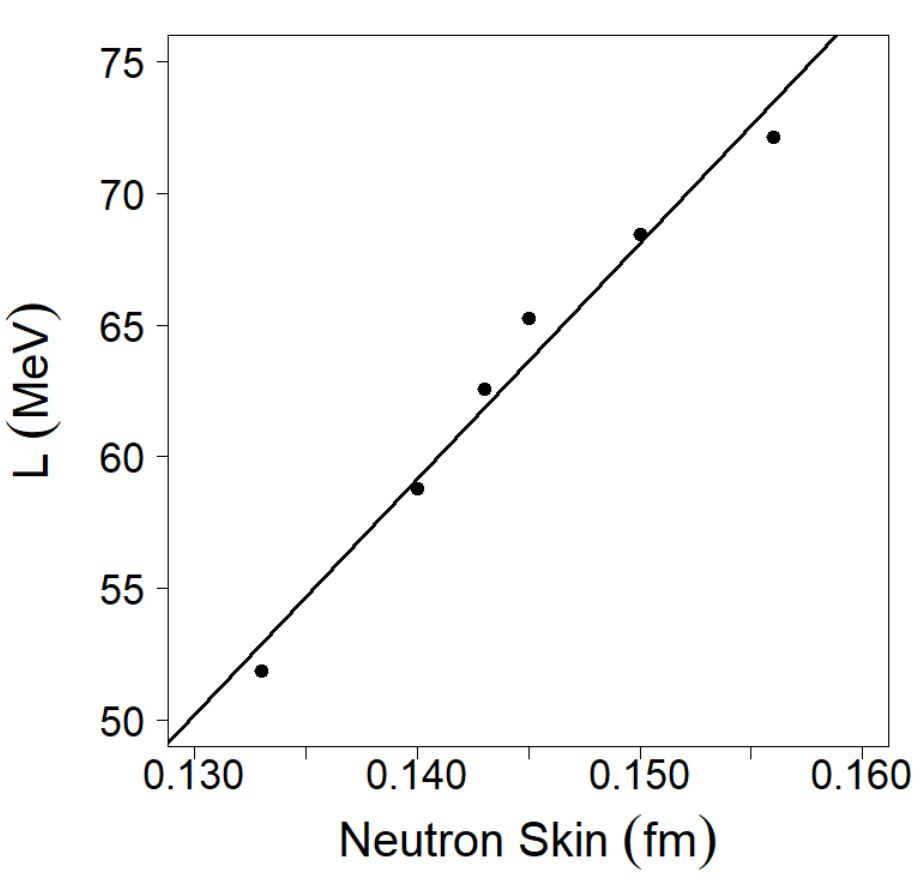}
\caption{Slope parameter (L) {\it vs.} neutron skin thickness of $^{208}$Pb. 
The correlation coefficient is equal to 0.988.}
\label{L_skin} 
\end{figure}

As shown in Fig.~\ref{L_skin}, we find an excellent correlation between the predicted neutron skin thickness of
$^{208}$Pb and the slope parameter. Table~\ref{ns_vals} shows the values of the neutron skin thickness predictions along with the truncation error. It is worth noting that a systematic improvement of the convergence pattern is best seen when the cutoff of 450 MeV is chosen. This is the case for the symmetry energy, the slope parameter and the neutron skin thickness of $^{208}$Pb. 
A similar study is shown in Table~\ref{ns_ca_vals} for $^{48}$Ca. 

Proceeding as previously, and taking the largest of the errors at the two cutoff values, 
we state our final estimates for the neutron skins of 
$^{208}$Pb and 
$^{48}$Ca as   
\begin{equation}
S_n(^{208}Pb) = 0.144 \pm \Delta_{S_n}      \; \; \; \; \;  \Delta_{S_n} = 0.009 (0.011) \; , 
\end{equation}  
\begin{equation}
S_n(^{48}Ca) = 0.140 \pm \Delta_{S_n}     \; \; \; \; \;  \Delta_{S_n} = 0.006 \; .          
\end{equation}  

\section{Summary and Conclusions}                                                                  
\label{Concl} 

We examined some important characteristics of the EoS which relate to the presence of isospin 
asymmetry, paying particular attention to the density dependence of the symmetry energy. 

First, we briefly reviewed how the 2NFs and the 3NFs employed here are developed.
We considered high-quality chiral 2NFs up to fifth order of the               
chiral perturbation expansion for two    
values of the cutoff parameter appearing in the regulator.      
As for the 3NF, we employ the leading chiral 3NF. However, for the 
2PE 3NF, as explained in Ref.~\cite{Sam2018}, we are able to include, effectively, contributions up to fifth order.   

We then moved on to the isospin-asymmetric EoS with a particular eye on the density derivatives of the 
symmetry energy. 
Overall, our microscopic chiral predictions, which
are presented along with their chiral truncation error, 
fall well within available empirical constraints.                                   

We also touched upon aspects of finite nuclei, such as the so-called
reference density, the asymmetry coefficient, and, finally, neutron skins. 
This was accomplished with the help of a simple liquid-drop inspired method which, in spite of its simplicity,
allows an immediate and direct connection with the input EoS. 
At this point, we took the opportunity to revisit the important correlation between the density slope of the 
symmetry energy and the neutron skin thickness of $^{208}$Pb, or, equivalently, the symmetry pressure 
and the neutron skin thickness.                                                            

Before closing, we like to point to our recent predictions for the radius of a typical-mass neutron 
star~\cite{Sam2019} obtained with the same chiral interactions as used in this paper and also well known to be
sensitive to the symmetry pressure. From that investigation, we 
obtained,                                
for the radius of a neutron star with a mass equal to 1.4 $M_{\odot}$ at N$^3$LO, an interval which 
is well within currently available constraints, see extensive list of citations in Ref.~\cite{Sam2019}. 

Our microscopic predictions of the EoS are obtained with high-quality chiral forces and considerations of theoretical uncertainties. They can be useful in guiding current and future               
empirical analyses of EoS-sensitive ``observables".

\section*{Acknowledgments}
Support by the U.S.\ Department of Energy, Office of Science, Office of Basic Energy Sciences, under Award Number DE-FG02-03ER41270 is acknowledged.

\end{document}